\documentclass[12pt]{article}

\def\mytitle#1{\setcounter{equation}{0}
\setcounter{footnote}{0}
\begin{flushleft}\Large\textbf{#1}\end{flushleft}
\vspace{0.27cm}}
\def\myname#1{\leftline{{\large #1}}\vspace{-0.13cm}}
\def\myplace#1#2{\small\begin{flushleft}\textit{#1}\\
\texttt{#2}\end{flushleft}}

\def\myclassification#1{\small\noindent
Pacs no : 04.20-q, 04.30-w, 04.40.Nr
       #1\vspace{0.5cm}}
\usepackage{graphicx}
\begin{document}

\mytitle{Gravitational collapse of dissipative fluid as a source of gravitational waves}

\vskip0.2cm \myname{Sanjukta Chakraborty\footnote{sanjuktachakraborty77 @gmail.com}}
\vskip0.2cm \myname{Subenoy Chakraborty\footnote{schakraborty@math.jdvu.ac.in}}

\myplace{Department of Mathematics, Tarakeswar Degree College, Tarakeswar, India.}{}
\myplace{Department of Mathematics, Jadavpur University, Kolkata-700 032, India.}{}

\begin{abstract}

Gravitational collapse of cylindrical anisotropic fluid  has been considered in analogy with the work of Misner and Sharp. Using Darmois matching conditions, the interior cylindrical dissipative fluid (in the form of shear viscosity and heat flux )is matched to an exterior  vacuum  Einstein--Rosen space-time. It is found that on the bounding $3$-surface the radial pressure of the anisotropic perfect fluid is linearly related to the shear viscosity and the heat flux of the dissipative fluid on the boundary. This non-zero radial pressure on the bounding surface  may be considered as the source of gravitational waves outside the collapsing matter distribution. \\

Keywords : Cylindrical collapse, Dissipation, heat flux, Junction conditions, Dynamical equations .
\end{abstract}
\myclassification{}\\
\section{Introduction}

In gravitational collapse, the final fate of a collapsing matter is a challenging issue in gravitational physics and particularly in relativistic astrophysics. During nuclear burning in a massive star, there is a stable configuration as the inward pull of gravity is balanced by the outward pressure of the nuclear fuel at the core of the star. However, when the nuclear fuel of the star has been exhausted due to thermonuclear burning the equilibrium configuration has been destroyed and there is a continual gravitational collapse.\\

The study of gravitational collapse was initiated long back in 1939 by Oppenheimer and Snyder [1] for homogeneous spherical dust cloud in general relativity. Subsequently, an inhomogeneous spherically symmetric dust cloud was analytically studied by Joshi and Singh [2] and they concluded that initial density distribution and the radius of the star plays a crucial role in characterizing the end state of collapse(i.e. black hole or naked singularity). Misner and Sharp[3] showed that pressure plays a crucial role at the end stages of collapse of a spherically symmetric ideal fluid with Schwarzschild space-time at the exterior of the star.\\

Although the majority of works on collapse dynamics deal with spherical model, still extensions to other kind of symmetries may provide important information about self-gravitating fluids in general. The natural non-spherical symmetry in nature is axial symmetry. Levi-Civita [4] first found the vacuum solution in Einstein gravity with cylindrical space-time and it still puzzles the relativists as the precise meaning of its two independent parameters is still unknown. Further, the basic features in the study of self gravitating compact objects having other kind of symmetries may provide important information about self gravitating fluids in general. In particular, there has been renewed interest in cylindrically symmetric sources in relation with different, classical and quantum, aspects of gravitation. Such sources may serve as test bed for numerical relativity, quantum gravity and for probing Cosmic censorship and hoop conjecture, among other important issues and represent a natural tool to seek the physics that lies behind the two independent parameters in Levi-Civita metric.  At first Herrera etal [5] showed that there is a non-vanishing radial pressure on the boundary surface, when a cylindrical non-dissipative fluid matches to an exterior containing gravitational waves. Then within a gap of two years the same authors with M.A.H. Maccallum [6] found the previous result to be wrong. Subsequently, Herrera and his collaborators considered cylindrical collapse of matter with[7] or without shear [8]. \\

Moreover, it should be noted that Darmois [9] junction conditions are the basic tools in studying gravitational collapse. By studying junction conditions between static exterior and non-static interior with a cosmological constant, Sharif and Ahmed [10-12] showed how collapsing process is slowed down due to the presence of the positive cosmological constant. In this context, Herrera etal [13] were able to show the mismatch between any conformally flat cylindrically symmetric static source and the Levi-Civita space-time by using Darmois junction conditions. Also, in the context of null dust collapse in the cylindrically symmetric space-time, Kurita and Nakao [14] showed that naked singularity is formed along the axis of symmetry.\\

On the otherhand, due to highly dissipative nature [15-17] of the collapsing process, it is more realistic to consider dissipative matter in this context. However, the dissipative process [15-17] is required to account for the very large(-ve) binding energy of the resulting compact object (of the order of $-10^{53}$). Further, the only plausible mechanism to carry away the bulk of the binding energy of the collapsing star, leading to a neutron star oa black hole is neutrino emission. Chan[18] has considered gravitational collapse of a radiating star with dissipation in the form of radial heat flow and shear viscosity. According to him, shear viscosity plays a significant role in the study of  gravitational collapse. The dynamical description of gravitational collapse with dissipation of energy as heat flow and radiation has been investigated by Herrera and Santos[16]. Considering casual transport equations related to different dissipative components (heat flow, radiation, shear and bulk viscosity) Herrera etal [13,19,20] studied the collapse dynamics. Recently, Sharif and Rehmat [21] has examined the same collapsing process with geometry in the form of plane symmetry.\\

In the seminal paper (section 6) on gravitational radiation of Bondi [22], it was clearly stated that not only in the case of dust, but also in the absence of dissipation in a perfect fluid, the system is not expected to radiate (gravitationally) due to the reversibility of the equation of state. One can argue thermodynamically as follows: radiation is an irreversible process, this fact emerges at once if absorption is taken into account and (or) Sommerfeld type conditions   (which eliminates inward travelling waves) are imposed. Hence, one needs an entropy generator factor(which is absent in a perfect fluid or in a collisionless dust) in the description of the source i.e. the irreversibility of the process of emission of gravitational waves, must be reflected in the equation of state through an entropy increasing factor (dissipative in nature). Recently, the link between the emission of gravitational radiation and the dissipative process has been described by  Herrera et al [23], for axially symmetric fluids.

In the context of gravitational waves, the sources must have symmetry different from spherical one. For bound sources, the exterior space-time should in principle describe such a radiation, but still now no exact solution in closed analytic form is not available to describe gravitational radiation from bound sources. Using null co-ordinates, only Bondi approach provides metric functions as inverse power series which converges very far from the source. In particular, no explicit exterior metric exists except in cylindrical symmetry one has Einstein-Rosen space-time.  Further, it is generally known in the context of gravitational waves that a radiating (gravitational waves) source always losses mass [22-25] as waves carry energy, but fluid pressure is unaffected. However, if one considers cylindrical  collapse of non-dissipative fluid with exterior containing gravitational waves, then by Darmois matching conditions, bounding surface must have non-zero pressure.\\

In the present work, we investigate cylindrical collapse of dissipative fluid with exterior vacuum in Einstein--Rosen co-ordinates [26]. Here the dissipation is in the form of shear viscosity and heat flux and by using Darmois matching conditions, it is examined how the dissipative effects  influences the collapse dynamics. The plan of the paper is as follows. The basic equations both for interior and exterior has been presented in section 2. Section 3 deals with the junction conditions on the boundary surface. Section 4 shows how the radial pressure on the boundary depends on the shear components, the metric components and heat flux by using the junction conditions. Solution to the emitted pulse has been evaluated in section 5. Finally there is summary of the work in the section 6. \\

\section{Interior matter distribution and basic equations}

Let us consider a collapsing cylindrical surface $\Sigma$ bounding an anisotropic dissipative fluid. So the time-like three surface $\Sigma$ divides four dimensional space-time into $M^-$ (interior) and $M^+$ (exterior) manifolds with $M^{-}\bigcap M^{+}=\phi$. Assuming co-moving coordinates inside hypersurface $\Sigma$, the line element for the interior region is given by
\begin{equation}
d{s_-^2}=-{A^2}d{t^2} +{B^2 }d{r^2} +{C^2}d{\phi^2} +{D^2}d{z^2}
\end{equation}\\
where A,B,C and D are functions of t and r.
Now to maintain cylindrical symmetry, we impose restriction on the coordinates as:\\
$-\infty\leq t\leq +\infty,~~~r\geq 0,~~~-\infty<z<+\infty,~~~0\leq \phi \leq 2\Pi$\\
and the hypersurfaces $\phi =0,  2\Pi$. Here $\xi_{z}=\delta_{z}$ and $\xi_{\phi}=\delta_{\phi}$ are the two killing vectors.
For cylindrically symmetry, one needs some physical and geometrical conditions to be imposed. In case of regular symmetry axis (to have gravitationa collapse) one must impose the following conditions[27]\\

(i) Existence of symmetry axis:
         $X\equiv\left|\xi^{\mu}_{\phi}\xi^{\nu}_{\phi}g_{\mu\nu}\right|\longrightarrow 0 ~~~as ~~r\longrightarrow 0^{+}$\\
				
				              i.e.  $C\longrightarrow 0~~ as~~ r\longrightarrow 0^{+}$
				where the radial coordinate is chosen such that $r=0$ is the position of the axis.\\
				
	(ii) Locally flatness near the symmetric axis:\\
				$\frac{X_{,\alpha}X_{,\beta}g^{\alpha\beta}}{4X}\longrightarrow~~ -1~~$ as $r\longrightarrow 0$\\
				i.e.$C^{'}\longrightarrow 0, \dot{C}\longrightarrow A$ as $r \longrightarrow 0$\\
				
(iii) Non- existence of closed time-like curves (CTC):\\

        $ \xi^{\mu}_{\phi}\xi^{\nu}_{\phi}g_{\mu\nu} < 0, i.e.\left\|\xi_{\phi}\right\|^{2}< 0$\\
				in the whole space -time.\\
				
(iv)  Asymptotic flatness: The space-time should be asymptotically flat in the radial direction i.e. $A,B,D \longrightarrow 1$ as $r\longrightarrow \infty$		 and $C\longrightarrow r$ as $r\longrightarrow\infty$	\\

The interior coordinate are notationally written as $\lbrace x^{-\mu} \rbrace \equiv[t,r,\phi,z] (\mu=0,1,2,3) $.\\

The energy momentum tensor for such a fluid which undergoes dissipation in the form of shear viscosity and heat flow is given by [6,7]

\begin{equation}
T_{\mu\nu}=(\rho+p_t){v_\mu}{v_\nu}+{p_t}g_{\mu\nu}+({p_r}-{p_t}){\chi_\mu}{\chi_\nu}-2\eta\sigma_{\mu\nu}+q_{\mu}v_{\nu}+q_{\nu}v_{\mu}
\end{equation}\\
where $\rho,~p_r,~p_t~ and~~\eta$ are the energy density ,the radial pressure, the tangential pressure and the coefficient of shear viscosity.
Here $v_\mu$ and $\chi_\mu$ and $q_{\mu}$ are  unit time-like, space-like and radial heat flux vector respectively satisfying
\begin{equation}
v_\mu v^\mu =-\chi_\mu\chi^\mu=-1~~~,~~~~\chi^\mu v_\mu=0~~~,~~~q_{\mu}v^{\mu}=0~~~~
\end{equation}\\
Moreover,  the shear tensor $\sigma_{\mu\nu}$ has the expression
\begin{equation}
\sigma_{\mu\nu}= v_{(\mu;\nu)}+a_{(\mu}v_{\nu)}-\frac{1}{3}\Theta(g_{\mu\nu}+v_\mu v_\nu)
\end{equation}\\
where $a_\mu=v_{\mu;\nu}v^\nu $ is the acceleration vector and $\Theta=v^\mu;_\mu$ is the expansion scalar. \\

Note that the term $\eta$ describing the shear viscosity in equation (2) corresponds to the standard irreversible thermodynamics (Eckart and Londau), which predicts the propagation of perturbations with infinite speed [20]. To overcome such difficulties, various relativistic theories with non-vanishing relaxation times have been proposed. However, due to its simple nature, the standard irreversible thermodynamics is still widely used.\\

 Let us choose in the above co-moving co-ordinates the four velocity,  the unit space-like vector and heat flux vector of the fluid as
\begin{equation}
v^\mu=A^{-1}\delta_0^\mu~~~,~~~~~~~~~~~~~~\chi^\mu=B^{-1}\delta_1^\mu~~~,~~~~~q^{\mu}=q\delta^{\mu}_{1}~~~~~
\end{equation}\\
Here the non-vanishing components of the shear tensor are
\begin{equation}
\sigma_{11}=\frac{B^2}{3A}[\Sigma_1-\Sigma_3]~~,~~~~~~\sigma_{22}=\frac{C^2}{3A}[\Sigma_2-\Sigma_1]~~~,\sigma_{33}=\frac{D^2}{3A}[\Sigma_3-\Sigma_2]~~~and~~~~~\sigma^2=\frac{1}{6A^2}[\Sigma_1^2+\Sigma_2^2+\Sigma_3^2]
\end{equation}\\
where
\begin{equation}
 \Sigma_1=\frac{\dot{B}}{B}-\frac{\dot{C}}{C}~~,~~~\Sigma_2=\frac{\dot{C}}{C}-\frac{\dot{D}}{D}~~,~~~\Sigma_3=\frac{\dot{D}}{D}-\frac{\dot{B}}{B}
\end{equation}\\
Also the explicit form of the acceleration vector and the expansion scalar are given by

\begin{equation}
a_1=\frac{A^\prime}{A}~~,\Theta=\frac{1}{A}(\frac{\dot{B}}{B}+\frac{\dot{C}}{C}+\frac{\dot{D}}{D})
\end{equation}\\
where, $\cdot$ $\equiv\frac{\partial}{\partial t}$~~~~~and~~~~~$^\prime$ $\equiv\frac{\partial}{\partial r}.$\\

The Einstein field equations $G_{\mu\nu}=\kappa T_{\mu\nu}$ for the metric (1) (using equations ((2)-(8))), reduces to five non-zero components, but we shall need only the following two:
$G_{11}=\kappa T_{11}$
i.e.\\
\begin{equation}
\frac{1}{A^2}[\frac{\ddot{C}}{C}+\frac{\ddot{D}}{D}+\frac{\dot{C}}{C}\frac{\dot{D}}{D}-\frac{\dot{A}}{A}\frac{\dot{C}}{C}-\frac{\dot{A}}{A}\frac{\dot{D}}{D}]-\frac{1}{B^2}[\frac{C^\prime}{C}\frac{D^\prime}{D}+\frac{A^\prime}{A}\frac{C^\prime}{C}+\frac{A^\prime}{A}\frac{D^\prime}{D}]=- \kappa [p_r-\frac{2\eta}{3A}(\Sigma_1-\Sigma_3)]
\end{equation}\\
and $G_{10}=\kappa T_{10}$ i.e.,
\begin{equation}
-\frac{\dot{C}^\prime}{C}-\frac{\dot{D}^\prime}{D}+\frac{C^\prime}{C}\frac{\dot{B}}{B}+\frac{D^\prime}{D}\frac{\dot{B}}{B}+\frac{A^\prime}{A}\frac{\dot{C}}{C}+\frac{A^\prime}{A}\frac{\dot{D}}{D}=\kappa AB^{2}q
\end{equation}\\
The non-zero components of the energy conservation relation $T_{\nu;\mu}^\mu=0$ are given by
\begin{equation}
\dot{\rho}+(\rho+p_r)\frac{\dot{B}}{B}+(\rho+p_t)(\frac{\dot{C}}{C}+\frac{\dot{D}}{D})+q^\prime A+(2\frac{A^\prime}{A}+\frac{B^\prime}{B}+\frac{C^\prime}{C})qA=0
\end{equation}
and
\begin{equation}
p_r^\prime+(\rho+p_r)\frac{B^\prime}{B}+(p_r-p_t)(\frac{C^\prime}{C}+\frac{D^\prime}{D})+\dot{q}\frac{A}{B^2}+(3\frac{\dot{B}}{B}+\frac{\dot{C}}{C})q\frac{B^2}{A}=0
\end{equation}\\

The exterior vacuum space-time $(M^{+})$ of the cylindrical surface $\Sigma$ is described by the metric in Einstein-Rosen co-ordinates [26] as
\begin{equation}
ds_+^2=-e^{2(\gamma-\psi)}(dT^2-dR^2)+e^{2\psi}dz^2+e^{-2\psi}R^2d\phi^2
\end{equation}

with $\psi=\psi(T,R)$,~~~~~~~~~~~$\gamma=\gamma(T,R)$~~~~~~~~~~and~~~notationally$x^{+\mu}=(T,R,\phi,z)$

The vacuum Einstein field equations $R_{\mu\nu}^+=0$ gives the cylindrically symmetric  wave equation in an Euclidean space-time i.e.
\begin{equation}
\psi_{,TT}-\psi_{,RR}-\frac{\psi_{,R}}{R}=0
\end{equation}
with
\begin{equation}
\gamma_{,T}=2R\psi_{,T}\psi_{,R}~~~~~,\gamma_{,R}=R(\psi_{,T}^2+\psi_{,R}^2)
\end{equation}

The exterior wave equation (14) is an indication for the possible existence of a gravitational wave field.

\section{Smooth matching of $M^{-}$ and $M^{+}$ across $\Sigma$ : Darmois junction conditions}

In this section we formulate the junction conditions for the interior manifold $(M^{-})$ and exterior manifold $(M^{+})$ across the bounding three surface $\Sigma$.
From the point of view of the interior manifold we can write down the surface $\Sigma$ as
\begin{equation}
f_{-}(t,r)=r-r_{\Sigma}=0
\end{equation}
where $r_\Sigma$ is a constant as $\Sigma$ is a comoving surface forming the boundary of the fluid. Using eq.(16) in (1) we have the interior space-time on $\Sigma$ as
\begin{equation}
ds_-^2\stackrel{\Sigma}{=}-d\tau^2 +C^2dz^2+D^2d\phi^2
\end{equation}
 where
\begin{equation}
 d\tau\stackrel{\Sigma}{=}Adt
\end{equation}
defines the time co-ordinate on $\Sigma$ and $\stackrel{\Sigma}{=}$ indicates the equality of both sides on the surface $\Sigma$.\\

We choose $\xi^0= \tau,\xi^2= z,~~\xi^3=\phi$ as the parameters on $\Sigma$ for our convenience.\\

Now from the point of view of the exterior manifold, we can write down the boundary surface $\Sigma$ as

\begin{equation}
f_{+}(T,R)=R-R_\Sigma(T)=0
\end{equation}
  Using (19) in (13), we have the exterior space-time on $\Sigma$ as
\begin{equation}
ds_+^2\stackrel{\Sigma}{=}-e^{2(\gamma-\psi)}	[1-(\frac{dR}{dT})^2]dT^2+e^{2\psi}dz^2+e^{-2\psi}R^2d\phi^2
\end{equation}
For smooth matching of the interior and exterior manifolds on the bounding three surface $\Sigma$(not a surface layer), Darmois conditions [9] can be stated as follows:-\\

The continuity of first fundamental form. This implies the continuity of the line element over the hypersurface.
\begin{equation}
(ds^2)_{\Sigma}=(ds^2_{-})_{\Sigma}=(ds^{2}_{+})_{\Sigma}
\end{equation}
The continuity of the second fundamental form i.e $K_{ij}d\xi^{i}d\xi^{j}$ implies the continuity of the extrinsic curvature $K_{ij}$ over the hypersurface [9] i.e.
 \begin{equation}
[K_{ij}]\equiv K_{ij}^+ -K_{ij}^-=0
\end{equation}

where $K_{ij}^\pm$ is given by

\begin{equation}
K_{ij}^\pm=-n_\sigma^\pm[\frac{\partial^2x_\pm ^\sigma}{\partial\xi^i \partial\xi^j}+\Gamma_{\mu\nu}^\sigma \frac{\partial x_\pm^\mu}{\partial\xi^i}\frac{\partial x_\pm^\nu}{\partial\xi^j}],~~~(\sigma,~\mu,~\nu~=0,1,2,3)
\end{equation}
Here $n_\sigma^\pm$ are the components of the outward unit normal to the hypersurface in the co-ordinates $x^{\pm\mu}$, and the christoffel symbols are calculated for the metric in $M^{-}$ or $M^{+}$ accordingly\\

The explicit components of the unit normal vectors are of the form

$n_\sigma^- \stackrel{\Sigma}{=}(0,B_\Sigma,0,0)~~and~~~~n_\sigma^+ \stackrel{\Sigma}{=}e^{2(\gamma-\psi)}(-R_\tau,T_\tau,0,0)$

where suffix $\tau$ indicates differentiation w.r.t. $\tau$.\\

So, the junction conditions due to continuity of the 1st fundamental form across $\Sigma$ gives
\begin{equation}
d\tau\stackrel{\Sigma}{=}e^{\gamma-\psi}[1-(\frac{dR}{dT})^2]^{1/2}dT\stackrel{\Sigma}{=}Adt
\end{equation}

\begin{equation}
C\stackrel{\Sigma}{=}e^\psi
\end{equation}

\begin{equation}
D\stackrel{\Sigma}{=}e^{-\psi}R
\end{equation}

Note that as T is time-like coordinate so $[1-(\frac{dR}{dT})^2]>0$ on $\Sigma$\\

The non vanishing components of extrinsic curvature in terms of interior and exterior coordinates are\\
\begin{equation}
 K_{00}^-\stackrel{\Sigma}{=}-\frac{A^\prime}{AB}
\end{equation}
\begin{equation}
K_{00}^+\stackrel{\Sigma}{=}e^{2(\gamma-\psi)}[T_{\tau\tau}R_\tau-R_{\tau\tau}T_\tau-(T_\tau^2-R_\tau^2)[R_\tau(\gamma_{,T}-\psi_{,T})+T_\tau(\gamma_{,R}-\psi_{,R})]]
\end{equation}\\
 \begin{equation}
K_{22}^-\stackrel{\Sigma}{=}\frac{C{C}^\prime}{B}
\end{equation}
\begin{equation}
K_{22}^+\stackrel{\Sigma}{=}e^{2\psi}(R_\tau \psi_{,T}+T_\tau \psi_{,R})
\end{equation}\\
and\\
\begin{equation}
K_{33}^-\stackrel{\Sigma}{=}\frac{-D{D}^\prime}{B}
\end{equation}
\begin{equation}
K_{33}^+\stackrel{\Sigma}{=}e^{-2\psi}R^2 (R_\tau \psi_{,T}+T_\tau \psi_{,R}-\frac{T_\tau}{R})
\end{equation}
Now the continuity of the second fundamental form, yields,
\begin{equation}
e^{2(\gamma-\psi)}[T_{\tau\tau}R_\tau-R_{\tau\tau}T_\tau-(T_\tau^2-R_\tau^2)[R_\tau(\gamma_{,T}-\psi_{,T})+T_\tau(\gamma_{,R}-\psi_{,R})]]=-\frac{A^\prime}{AB}
\end{equation}\\
\begin{equation}
e^{2\psi}(R_\tau \psi_{,T}+T_\tau \psi_{,R})=\frac{CC^\prime}{B}
\end{equation}\\
\begin{equation}
e^{-2\psi}R^2 (R_\tau \psi_{,T}+T_\tau \psi_{,R}-\frac{T_\tau}{R})=\frac{-DD^\prime}{B}
\end{equation}\\

\section{Results using matching conditions and field equations: Non zero radial pressure on the boundary}
In this section, we shall write down the matching conditions (formulated in the previous section) in a concise form by using the interior and exterior field equations.
From equation (24) we write
\begin{equation}
T_\tau^2 -R_\tau^2\stackrel{\Sigma}{=}e^{2(\psi-\gamma)}		
\end{equation}\\
Using the matching conditions (25) and (26) in (18) we have
\begin{equation}
R_\tau\stackrel{\Sigma}{=}\frac{(CD),t}{A}
\end{equation}\\
Also from the continuity of $K_{22}$ and $K_{33}$ with (25) and (26) we obtain
\begin{equation}
T_\tau\stackrel{\Sigma}{=}\frac{1}{B}(CD),r
\end{equation}\\
Now differentiating (37) and (38) with respect to $\tau$ and using the Einstein field equations (9) and (10) we get

\begin{eqnarray}
 T_{\tau\tau}R_\tau-T_\tau R_{\tau\tau}&=&\frac{1}{A^3 B}{A}^\prime(\dot{C}D+C\dot{D})^2+\frac{1}{A^2 B}(D{D}^\prime\dot{C}^2+C{C}^\prime\dot{D}^2)+\frac{\kappa qCDB}{A} (\dot{C}D+C\dot{D})+\frac{CD}{B}\kappa({C}^\prime D
\nonumber
\\
&+&C{D}^\prime)[p_r-\frac{2\eta}{3A}(\Sigma_1 -\Sigma_3)]-\frac{1}{B^3}({C}^\prime D+C{D}^\prime)({C}^\prime{D}^\prime+\frac{{A}^\prime{C}^\prime D}{A}+\frac{{A}^\prime{D}^\prime C}{A})
\end{eqnarray}\\

Differentiating (25) and (26) w.r.t $\tau$  and using the continuity of $K_{22}$ and $K_{33}$ we get [29],

\begin{equation}
\psi_{,T}(T_{\tau} ^2 -R_{\tau} ^2)=\frac{1}{AB}(\dot{C} D^\prime- C^ \prime \dot{D})
\end{equation}

Substituting (39)and (40) in the continuity of $K_{00}$  and using the field equations (15) we obtain (after a bit algebra) [29]

\begin{equation}
 \frac{(CD)_{,r}}{B^2}\kappa[p_r -\frac{2\eta}{3A}(\Sigma_1-\Sigma_3)]+\frac{q(CD)_{,t}}{A}\stackrel{\Sigma} {=} 0
\end{equation}\\

i.e. $p_{eff}\equiv\kappa[p_r -\frac{2\eta}{3A}(\Sigma_1-\Sigma_3)]+q\frac{B^2(CD)_{,t}}{A(CD)_{,r}}=0$~~~~(provided $(CD)_{,r}\neq0$)\\

The above result shows that the radial pressure $p_r$ on the boundary surface $\Sigma$ does not vanish rather depends linearly on shear viscosity and heat flux. However,  in absence of dissipation, radial pressure vanishes on the boundary and it agrees with the result of Herrera etal [6]. Moreover, from equation(41) we have the following observations.\\

(a)If the product of the metric coefficients $C$ and $D$ is constant (i.e.$CD=\lambda,$ a constant) then equation (41) is trivially satisfied.\\

(b)If the product $'CD'$ is independent of time then the radial pressure on the boundary is independent of heat flux while for independence of $CD$ over radial co-ordinate 'r' demands either $q=0$ or $CD$ should be constant.\\

(c)If the metric co-efficient 'B' is the g.m of the other two metric co-efficients $C$ and $D$ $(i.e. B^{2}=CD)$ then $p_{r}$ depends only on the dissipative heat flow, while if metric co-efficients $B$ and $CD$ are independent of temporal co-ordinate then $p_r$ is identically zero on $\Sigma$ although there is dissipation .\\

(d) From the symmetry conditions we see that as $r\longrightarrow 0^{+}$, $C^\prime \longrightarrow 0$ and $\dot{C}\longrightarrow A$. So from equation (41) we have  $q\longrightarrow 0$ as $r\longrightarrow 0^{+}$\\

Further, if the exterior space-time is chosen to be static (i.e. $\psi=\psi(R)$) then from equations (25), (26)  and (37) we have $C\stackrel{\Sigma}{=}C(r)$ and $D\stackrel{\Sigma}{=}D(r)$ i.e. $R\stackrel{\Sigma}{=}$ constant. So there will no longer be any collapse. However, from eq.(41) the radial pressure on the boundary is still non-zero due to the presence of shear viscosity. Hence no matching is possible and is consistent with the known result of mismatch of Levi-civita space-time (i.e.static exterior space-time) with a collapsing cylinder with matter source (for dust source this result has been shown in [29]). \\

\section{Cylindrical waves and its properties}
The monochromatic wave as solution of the cylindrical wave equation (14) can be written as [30]\\
\begin{equation}
\psi(T,R)=c_{0}J_{0}({\omega}R)cos({\omega}T+\alpha_{0})+d_{0}N_{0}({\omega}R)cos({\omega}T+\beta_{0})
\end {equation}

where $J_{0}$ and $N_{0}$ are the usual zeroth order Bessel functions of the first and second kinds respectively, $\omega$ is the angular frequency of the monochromatic wave and $c_{0}$, $d_{0}$, $\alpha_{0}$ and $\beta_{0}$ are arbitrary constants. In particular, for a standing wave of the form
\begin{equation}
\psi=c_{0}J_{0}({\omega}R)cos({\omega}T)
\end {equation}
One can solve for $\gamma$ from equation(15) as
\begin{equation}
\gamma(T,R)=\frac{1}{2}c_{0}^{2}{\omega}RJ_{0}({\omega}R)J_{0}^{'}({\omega}R)cos({2\omega}T)+\frac{1}{2}c_{0}^{2}{\omega}^{2}R^{2}[[J_{0}^{'}({\omega}R)]^{2}-J_{0}({\omega}R)J_{0}^{''}({\omega}R)]
\end {equation}

Note that both $\psi$ and $\gamma$ are singularity free periodic function of $T$ and these standing waves are developed due to reflection at the surface of a large sphere having centre at the origin. On the otherhand, if we replace $J_{0}$ by $N_{0}$ in equations (43) and (44) the wave solution has a singularity at the origin and it can be interpreted as a standing cylindrical gravitational wave with matter along the z-axis.\\

Further, for outgoing wave if we choose the solution as
\begin{equation}
\psi(T,R)=c[J_{0}({\omega}R)cos({\omega}T)+N_{0}({\omega}R)sin({\omega}T)]
\end {equation}
then in the asymptotic limit (i.e. for large $R$) $\psi$ takes the form

\begin{equation}
\psi{\approx}c(\frac{2}{\Pi\omega R})^{1/2}cos({\omega}R-{\omega}t-\frac{\Pi}{4})
\end {equation}

 Such a wave carries away energy from the matter located along the z-axis and as a result there should be a change in the motion, causing $\psi$ to be aperiodic. Hence such a solution is not of much physical interest.\\
       We shall now consider pulse solutions for cylindrical wave. Suppose a cylindrical wave starts at the z-axis in the form of a disturbance of short duration and travels outward from the axis. Then $\psi$ can be written in an integral form as [30]
\begin{equation}
\psi(T,R)=\frac{1}{2\Pi}\int^{\tau}_{-\infty}\frac{f(T^\prime)dT^\prime}{[(T-T^\prime)^2-R^2]^\frac{1}{2}}+\psi_{C}
\end{equation}

where $\tau=T-R$ is the usual retarded time, $f(T)$ indicates the strength of the wave source and $\psi_{C}$ is the standard Levi -Civita static solution [4]. \\
We shall now consider the following choices for the function $f(T)$.\\

I. Sharp pulse: $f(T)=f_{0}\delta(T)$
Here $f_{0}$ is a constant and $\delta(T)$ is the Dirac delta function. Then $\psi$ takes the form

\begin{equation}
\psi=\left\{
\begin{array}{ll}
\psi_{C} & \mbox{ for } \tau{< }0\\
\frac{f_0}{2\Pi(T^2-R^2)^{\frac{1}{2}}}+\psi_{C} & \mbox{ for } \tau{>}0
\end{array}
\right\}
\end{equation}
 The function $\psi$ and its derivative are regular everywhere except at the wave front determined by the surface at which the retarded time $\tau=T-R$, followed by a tail which persists for a long time. \\
II. The source function $f(T)$ is continuous of the form
\begin{equation}
f(T)=\left\{
\begin{array}{ll}
0 & \mbox{ for } T{< }0\\
f_{0}T & \mbox{ for } T{>}0
\end{array}
\right\}
\end{equation}
Then $\psi$ takes the form

\begin{equation}
\psi(T)=\left\{
\begin{array}{ll}
\psi_{C} & \mbox{ for } \tau{< }0\\
\psi_{C}+\frac{f_0}{2\Pi}[T\ln [\frac{T+\sqrt{T^2-R^2}}{R}]-(T^2-R^2)^{\frac{1}{2}}] & \mbox{ for } \tau{>}0
\end{array}
\right\}
\end{equation}
which is well behaved at $\tau=0$.\\

III. Source function  of finite duration:
\begin{equation}
f(T)=\left\{
\begin{array}{ll}
0 & \mbox{ for } T{< }0\\
0 & \mbox{ for } T{>}T_{c}
\end{array}
\right\}
\end{equation}

In this case if we take $\tau=T-R>>T_{c}$ then the integral of the equation (47) can be approximated as
\begin {equation}
\psi\approx\psi_{C}+\frac{1}{2\Pi}\frac{f_{0}}{(T^2-R^2)^{\frac{1}{2}}}
\end{equation}

where  the constant $f_{0}$ is given by
$f_{0}=\int_{0}^{T_{c}}f(T^{'})dT^{'}$

In this case also the 'tail' of the wave is free from singularities. Hence it is possible to have well behaved $\psi$ considering a continuous source function of finite duration. However, generation of such wave is still an open question. Further, source function can be refereed to as a mathematical device for obtaining solution of the field equations with a given behaviour near the z-axis. Finally, it should be kept in mind that from physical point of view, the z-axis should be replaced by a tube of small but finite cross section for the given distribution of matter so that the source function would be expressed in terms of integrals involving the energy- momentum tensor and consequently the waves emitted would have to be of a form compatible with these conditions.\\

\section{Short Discussion:}
Cylindrical collapse of dissipative fluid has been studied using the formulation of Misner and Sharp. Here both shear viscosity sand heat flux are chosen as the dissipative character. The collapse dynamics is studied using matching conditions on the boundary. Following Darmois, the junction conditions are evaluated on the boundary having exterior manifold as vacuum cylindrically symmetric space-time due to Einstein and Rosen. It is found from the junction conditions that the radial pressure is non-zero on the bounding surface due to the dissipative nature of the collapsing matter. From the linear relation between the dissipative factors (namely shear viscosity and heat flux) and the radial pressure on the boundary, we have shown various situations depending on the functional dependence of the product of the metric co-efficients $'C'$ and $'D'$. However, the vanishing of the effective radial pressure on the boundary is consistent with the equivalence principle and it agrees with the result of Herrera etal [6] and our earlier work [28]. In fact, vanishing of the radial pressure on the boundary surface implies that any local pressure effect, predicted by a local characterization of the energy (momentum) flux of gravitational energy of the Einstein-Rosen waves, by means of pseudo-tensors [31], will not be observed. A detailed study of the cylindrical wave is presented in sec.$5$. It is found that monochromatic waves are not of much physical interest in the present context. However, pulse wave with continuous source function of finite duration is interesting in the present context. More over, the present work is consistent with the result of Bondi [32]  that there will be no longer any radiation of gravitational waves if we have collapsing dust cylinder. Finally, we may conclude that dissipative collapsing cylindrical matter may generate gravitational waves to exterior non-static vacuum space-time.\\

\section{Acknowledgement}
One of the author (SC) is thankful to UGC-DRS programme, Department of Mathematics, Jadavpur University.

\end{document}